\documentclass[12pt,reqno]{article}

\usepackage{mathtools}
\usepackage{dirtytalk}
\usepackage{amsmath}
\usepackage{amsfonts}
\usepackage{amssymb}
\usepackage{amsxtra}
\usepackage{amsthm}
\usepackage{graphicx}
\usepackage{caption}
\usepackage{hyperref}
\usepackage{ushort}
\usepackage{color}
\usepackage[parsep]{collref}
\hypersetup{linktocpage=true}
\usepackage{enumitem}
\usepackage{titlesec}
\usepackage{cite}
\usepackage{slashed}
\usepackage{braket}
\usepackage{flexisym}
\usepackage{xcolor}
\usepackage{dsfont}
\usepackage{titlesec}
\usepackage{romannum}
\usepackage{caption}
\usepackage{subcaption}
\usepackage{mathrsfs}
\usepackage[bottom]{footmisc}
\usepackage[bbgreekl]{mathbbol}
\usepackage{bbold}

\DeclareSymbolFontAlphabet{\mathbb}{AMSb}
\DeclareSymbolFontAlphabet{\mathbbl}{bbold}
\def\ie{{\it i.e.\ }}

\newcommand{\mb}{\mathbb}
\newcommand{\mc}{\mathcal}

\newcommand{\ov}{\overline}
\newcommand{\ph}{\phantom}

\DeclareMathOperator{\vol}{vol}

\DeclareMathOperator{\tr}{tr}
\DeclareMathOperator{\diag}{diag}

\newcommand{\subsubsubsection}[1]{\paragraph{#1}\mbox{}\\}
\newcommand{\imperial}{\it 
The Blackett Laboratory, Imperial College London\\
Prince Consort Road, London, SW7 2AZ
}
\newcommand{\auth}{Rahim Leung\,\footnote{\,rahim.leung14@imperial.ac.uk} and K.S. Stelle\,\footnote{\,k.stelle@imperial.ac.uk}}

\let\oldabstract\abstract
\let\oldendabstract\endabstract
\makeatletter
\renewenvironment{abstract}
{%
               {\list{}{\addtolength{\leftmargin}{1em} 
                        \listparindent 1.5em%
                        \itemindent    \listparindent%
                        \rightmargin   \leftmargin%
                        \parsep        \z@ \@plus\p@}%
                \item\relax}%
               {\endlist}%
\oldabstract}
{\oldendabstract}
\makeatother

\numberwithin{equation}{section}

\let\oldsection\section
\renewcommand{\section}{\renewcommand{\theequation}{\thesection.\arabic{equation}}\oldsection}

\titlespacing*{\section}
{0pt}{5.5ex plus 1ex minus .2ex}{4.3ex plus .2ex}
\titlespacing*{\subsection}
{0pt}{5.5ex plus 1ex minus .2ex}{4.3ex plus .2ex}

\begin{document}
\setcounter{page}{0}
\thispagestyle{empty}
\begin{flushright}
\hfill{
Imperial/TP/2022/KS/02}\\
\end{flushright} 
\vspace{20pt}

\begin{center}  

{\Large {\bf Supergravities on Branes}}   

\vspace{25pt}

\auth

\vspace{7pt}
\imperial

\end{center} 

\vspace{20pt}

\begin{abstract}

Supergravity brane solutions allow for a generalised type of Kaluza-Klein reduction onto brane worldvolumes. The known replacement of a flat worldvolume metric by a Ricci-flat metric constitutes a consistent Kaluza-Klein truncation of the starting higher-dimensional supergravity theory down to a lower-dimensional pure gravity theory. 



This paper shows how to extend such a brane-worldvolume pure-gravity consistent truncation to that for a full nonlinear supergravity theory for the Type IIB D3-brane and the M-theory/Type IIA M5-brane. The extension of worldvolume supersymmetry is given by the unbroken supersymmetry of the original flat \say{skeleton} brane. Compatibility with further worldvolume diagonal and transverse vertical dimensional reductions is also shown, providing the brane-worldvolume supergravity embeddings of all descendants of the skeleton D3- and M5-branes. Examples are given of brane-worldvolume supergravity solutions embedded into the corresponding higher-dimensional supergravities.

\end{abstract}
\vfill\leftline{}\vfill
\pagebreak

\tableofcontents
\addtocontents{toc}{\protect\setcounter{tocdepth}{2}}
\pagenumbering{arabic}
\setcounter{page}{1}
\setcounter{footnote}{0}

\section{Introduction}

The idea of a gravitational theory belonging to a Minkowski-signature subspace of a higher dimensional spacetime, \ie a braneworld gravity theory, has a long history in theoretical physics investigations, and has clearly been of interest for cosmology \cite{Rubakov:1983bb}. In Ref.\cite{Erickson:2021psj}, a categorisation, or ``taxonomy'' of such braneworld gravities was given for constructions involving noncompact transverse spaces. The conceptually simplest of these categories, \say{Type I}, involves a consistent truncation of the higher-dimensional supergravity to a lower-dimensional theory on  the worldvolume. This type will be the main focus of the present paper. It should be contrasted with the generation, not via a consistent truncation, of a lower-dimensional effective theory that is only valid within a certain low-energy range of scales, called \say{Type III} in Ref.\cite{Erickson:2021psj}. Examples of the latter include Randall-Sundrum (RS) models \cite{Randall:1999vf} involving patching of sections of a higher-dimensional AdS space, or models \cite{Crampton:2014hia} which naturally generate a mass gap between a discrete $L^2$ normalisable transverse-space wave function mode and the continuum of delta-function normalisable modes expected from a noncompact transverse space \cite{Hull:1988jw}. 

Type I and Type III constructions have different virtues. Type I constructions, involving a consistent truncation of the higher dimensional theory, give rise to fully nonlinear lower-dimensional worldvolume theories. Type III constructions, not based on a consistent truncation, inherently give only a perturbative realisation of lower-dimensional gravitational physics, but they do generate an intrinsic localisation of the effective gravity theory at least within a limited range of scales \cite{Erickson:2021psj,Erickson:2022qhv}.

The existence of Type I consistent truncations to worldvolume gravities was found in Ref.\cite{Brecher:1999xf} for domain-wall or more general magnetic braneworlds. These constructions promote the flat Minkowskian worldvolume metric of an initial \say{vacuum} supergravity brane solution to a generalisation with a worldvolume Ricci-flat metric. Within the context of the original higher dimensional theory, this amounts to making a consistent truncation to the lower-dimensional theory, but just for the worldvolume metric. This gave rise to applications such as braneworld black holes \cite{Chamblin:1999by}.

There can be a lot more to worldvolume physics than just Ricci-flat pure gravity metrics, however. An indication of this possibility was given in a study \cite{Lu:2000xc} of the embedding of Maxwell-Einstein $d=4$ supergravity within a $D=5$ RS gauged supergravity context, which was then lifted further up to $D=10$ Type IIB supergravity using an existing Kaluza-Klein $S^5$ ans\"atze \cite{Lu:1999bw}. In that case, the unbroken supersymmetry of the vacuum RS construction allowed an extension of a $d=4$ metric-only truncation to one involving also a lower-dimensional Maxwell field. This provided a good example of a Kaluza-Klein reduction down to a $d=4$ interacting gravity--matter theory with nontrivial dependence on the transverse space -- in that case, reduction on the single coordinate transverse to the RS patch surface.

Since an $\hbox{AdS}_5\times S^5$ spacetime is the asymptotic near-horizon geometry of a Type IIB D3-brane, the construction of Ref.\cite{Lu:2000xc} may be taken to suggest a further generalised embedding of D3 worldvolume (\ie horizon) supergravity into $D=10$ Type IIB supergravity. The original D3-brane is a $\frac12$-BPS solution, \ie it possesses sixteen unbroken supercharges. So the suggested worldvolume supergravity theory in the D3 brane case is $\mc{N}=4$, $d=4$ supergravity. The original flat Type IIB D3-brane solution should serve as a structural \say{skeleton} for such a construction, generalising the way that $D=5$ AdS space serves in the construction of Ref.\cite{Lu:2000xc}.

Showing how a fully nonlinear $\mc{N}=4$ supergravity theory embedding on a D3-brane worldvolume works is the first result of the present paper, as presented in Section \ref{sec:D3}. The resulting KK ans\"atze is motivated by considering carefully the two asymptotic limits: flat $D=10$ space at transverse-space infinity and $\hbox{AdS}_5\times S^5$ in the near-horizon region, with the result confirmed by showing agreement, after a voluntary truncation to just $\mc{N}=2$ supersymmetry, with the $d=4$ structure of Ref.\cite{Lu:2000xc}. We follow this up by demonstrating how a wide class of $\mc{N}=4$ worldvolume black-hole solutions is embedded into the Type IIB theory, using the harmonic-map construction of Refs \cite{Neugebauer:1969wr,Breitenlohner:1987dg,Clement:1996nh,Galtsov:1998mhf}.

In Section \ref{sec:M5}, we present an analogous construction for the M5-brane, with fully nonlinear $\mc{N}=(4,0)$, $d=6$ worldvolume supergravity. An example of an embedded solution into $D=11$ M-theory supergravity in this case is given by a worldvolume multi-charged anti-self-dual string solution.

Section \ref{sec:diagvert} shows how further diagonal and vertical dimensional reductions may consistently be combined with the braneworld embeddings, giving the braneworld supergravity embeddings of all descendants of the D3- and M5-branes.

\section{D3-branes with their worldvolume supergravities}\label{sec:D3}

The geometry of the $\frac12$-BPS solution of $D=10$ Type IIB supergravity describing parallel D3-branes is a warped product $\mb{R}^{1,3}\times_W\mb{R}^6$. As shown in Ref.\cite{Brecher:1999xf}, it is consistent to replace the $\mb{R}^{1,3}$ worldvolume of the D3-branes with a generic, four-dimensional manifold $M_4$ with a general Ricci-flat metric $g_{\mu\nu}$. In the language of Ref.\cite{Erickson:2021psj}, this is a \say{Type I} consistent truncation of Type IIB supergravity down to pure, $\mc{N}=1$, $d=4$ supergravity. However, since the original flat-worldvolume solution preserves sixteen supercharges, we anticipate the existence of a more general consistent truncation to pure $\mc{N}=4$, $d=4$ supergravity. The Ricci-flat replacement $\mb{R}^{1,3}\to M_4$ would then be just a further consistent truncation of the $\mc{N}=4$ theory to its purely gravitational sector. Our goal here is to construct such a generalisation of flat D3-branes, referred to henceforth as skeleton branes, to ones with full four-dimensional worldvolume $\mc{N}=4$ supergravity. We will then classify the stationary black-hole solutions of the $\mc{N}=4$ theory and, for completeness, will explicitly construct a family of charged black hole solutions. These solutions, when uplifted back into Type IIB supergravity using our embedding, have interpretations as D3-branes with charged black holes on their worldvolume.

\subsection{Skeleton D3-branes and $d=4$, $\mc{N}=4$ supergravity}

The supergravity solution of $D=10$ Type IIB supergravity describing $N$ parallel D3-branes in units where $\alpha'=1$ is given by\footnote{Our convention for the Hodge dual is $\ast(dx^{m_1}\wedge\cdots\wedge dx^{m_p}) = \frac{1}{q!}\sqrt{|g|}\epsilon_{\ph{m_1\cdots m_p}n_1\cdots n_q}^{m_1\cdots m_p} dx^{n_1}\wedge\cdots\wedge dx^{n_q}$, where $\epsilon_{n_1\cdots n_D}$ with lowered indices is numerical with $\epsilon_{012\dots D-1} = 1$, and $q = D-p$.}
\begin{equation}\label{D3branesol}
\begin{split}
&d\hat s^2 = H^{-1/2}ds^2(\mb{R}^{1,3}) + H^{1/2}\left(dr^2+r^2d\Omega^2_5\right) \,,\\
& \hat F_{(5)} = 16\pi N(1+\hat{\ast})\vol_5 \,,\quad H = 1+ \frac{4\pi N}{r^4} \,,
\end{split}
\end{equation}
where $d\Omega^2_5$ is the metric on the round five-sphere, $\vol_5$ is its associated volume form, and for simplicity, the string coupling constant has been set to one. This solution preserves sixteen rigid supercharges. For this to be a string theory background, the five-form flux must obey the quantisation condition
\begin{equation}
\frac{1}{16\pi^4}\int_{C_5}\hat F_{(5)} \in\mb{Z} \,,
\end{equation}
for any non-trivial five-cycle $C_5$. The only non-trivial five-cycle in \eqref{D3branesol} is $C_5= S^5$, and for it we find 
\begin{equation}
\frac{1}{16\pi^4}\int_{C_5}\hat F_{(5)} = N \in\mb{Z} \,.
\end{equation}
In terms of the Cartesian coordinates $y^\Lambda$, $\Lambda\in\{1,\dots,6\}$, on the transverse space $\mb{R}^6$, the volume form of the five-sphere is
\begin{equation}
\vol_5 = \frac{1}{5!}\frac{y^\Lambda}{r^6}\epsilon^\Lambda_{\ph{\Lambda}\Sigma_1\cdots\Sigma_5}dy^{\Sigma_1}\wedge\cdots\wedge dy^{\Sigma_5} = -\frac{1}{16\pi N}{\ast}_6dH \,,
\end{equation}
where $r^2 = y^\Lambda y^\Lambda$, $\epsilon_{123456}=1$, the $\Lambda$ indices are raised/lowered by the flat Euclidean metric $\delta_{\Lambda\Sigma}$, and ${\ast}_6$ is the Hodge dual with respect to $\delta_{\Lambda\Sigma}$. The five-form is then explicitly
\begin{equation}
\hat F_{(5)} = H^{-2}\vol_4\wedge dH - {\ast}_6dH \,,
\end{equation}
where $\vol_4$ is the volume form on the Minkowski worldvolume. 

The solution \eqref{D3branesol} is the skeleton on which we will embed the pure $\mc{N}=4$, $d=4$ supergravity theory. This theory contains one graviton, four gravitini, six $U(1)$ vectors, four dilatini, and one complex scalar, which parametrises the coset space $SL(2,\mb{R})/SO(2)$. We will denote the bosonic degrees of freedom by the metric $g_{\mu\nu}$, six one-forms $\mc{A}^\Lambda_{(1)}$, where $\Lambda\in\{1,\dots,6\}$, and the complex scalar $\tau = \chi + ie^{-\phi}$. The Lagrangian describing the bosonic sector of the $d=4$ theory is given by
\begin{equation}\label{goaltheory}
\mc{L}_4 = R{\ast}_41  - \frac{1}{2}d\phi\wedge{\ast}_4d\phi - \frac{1}{2}e^{2\phi}d\chi\wedge{\ast}_4d\chi-\frac{1}{2}e^{-\phi}\mc{F}^\Lambda_{(2)}\wedge{\ast}_4\mc{F}^\Lambda_{(2)} - \frac{1}{2}\chi\mc{F}^\Lambda_{(2)}\wedge\mc{F}^\Lambda_{(2)}\,.
\end{equation}
where $\mc{F}^\Lambda_{(2)} = d\mc{A}^\Lambda_{(1)}$. For the rest of this paper, all fields without a hat are taken to be lower-dimensional.

For completeness, we will record below the Bianchi identities and the equations of motion obtained from \eqref{goaltheory}:
\begin{eqnarray}
&&d\mc{F}^\Lambda_{(2)} = 0 \,,\label{4dbianchi} \\
&&d\mc{G}^\Lambda_{(2)} = 0\,, \label{eom4dvec} \\
&&d{\ast}_4d\phi - e^{2\phi}d\chi\wedge{\ast}_4d\chi + \frac{1}{2}e^{-\phi}\mc{F}^\Lambda_{(2)}\wedge{\ast}_4\mc{F}^\Lambda_{(2)} = 0 \,,\label{eom4dphi} \\
&&d(e^{2\phi}{\ast}_4d\chi) - \frac{1}{2}\mc{F}^\Lambda_{(2)}\wedge\mc{F}^\Lambda_{(2)} = 0\,,\label{eom4dchi} \\
&& R_{\mu\nu} = \frac{1}{2}\nabla_\mu\phi\nabla_\mu\phi + \frac{1}{2}e^{2\phi}\nabla_\mu\chi\nabla_\nu\chi + \frac{1}{2}e^{-\phi}\left(\mc{F}^\Lambda_{\mu\rho}\mc{F}^{\Lambda\,\rho}_\nu - \frac{1}{4}(\mc{F}^\Lambda)^2g_{\mu\nu}\right) \,,\label{4dEinstein}
\end{eqnarray}
where $(\mc{F}^\Lambda_{(2)})^2 = \mc{F}^\Lambda_{\mu\nu}\mc{F}^{\Lambda\,\mu\nu}$, and 
\begin{equation}\label{sl2dual4d}
\mc{G}^\Lambda_{(2)} = e^{-\phi}{\ast}_4\mc{F}^\Lambda_{(2)} + \chi\mc{F}^\Lambda_{(2)} 
\end{equation}
is the $SL(2,\mb{R})$ dual of $\mc{F}^\Lambda_{(2)}$.

\subsection{Embedding $\mc{N}=4$ supergravity on D3-branes}

To embed the pure $\mc{N}=4$ supergravity theory on the worldvolume of the skeleton D3-branes, we need to generate six Abelian vector fields and a complex scalar that parameterises the coset $SL(2,\mb{R})/SO(2)$. These vector fields cannot be obtained through a usual Kaluza-Klein reduction on the transverse $\mb{R}^6$, as the D3-branes are supported by a harmonic function on that $\mb{R}^6$. They also cannot be generated through a Pauli reduction on the transverse $S^5$, \ie the five-sphere boundary of $\mb{R}^6$ at infinity, because $U(1)^6$ is not a subgroup of $SO(6)$. This means that the six vector fields must appear in the fluxes of the Type IIB theory, and, due to gauge invariance, they must appear through their field strengths. The field strengths must also appear in an $SO(6)$ invariant manner, as the $\mc{N}=4$ theory has a global $SO(6)$ symmetry that rotates them. 
Furthermore, the $D=10$ Type IIB theory already contains a complex scalar $\hat\tau = \hat C_0 + ie^{-\hat\Phi}$ that parameterises the $SL(2,\mb{R})/SO(2)$ coset. With these facts in mind, we find, after some trial and error, that the embedding of the full $d=4$, $\mc{N}=4$ supergravity theory on the worldvolume of skeleton D3-branes is given by
\begin{equation}\label{N=4ansatz}
\begin{split}
&d\hat s^2 = H^{-1/2}g_{\mu\nu}(x)dx^\mu dx^\nu + H^{1/2}d{y}^\Lambda d{y}^\Lambda \,,\quad \hat\Phi= \phi(x)\,,\quad \hat C_0 = -\chi(x) \,,\\
&\hat H_{(3)} = \frac{1}{\sqrt{2}}\mc{F}^\Lambda_{(2)}\wedge d{y}^\Lambda \,,\quad \hat F_{(3)} = -\frac{1}{\sqrt{2}}e^{-\phi}{\ast}_4\mc{F}^\Lambda_{(2)}\wedge d{y}^\Lambda \,, \\
&\hat F_{(5)} = H^{-2}\vol_4\wedge dH - {\ast}_6dH \,,\quad H = 1 + \frac{4\pi N}{r^4} \,.
\end{split}
\end{equation}
where $\mc{F}^\Lambda_{(2)}$ are two-forms on the four-dimensional subspace that we will show correspond to the field strengths of the $\mc{N}=4$ theory, ${\ast}_4$ is the Hodge dual computed with respect to the four-dimensional metric $g_{\mu\nu}$, $\vol_4$ is the volume form associated with $g_{\mu\nu}$, and ${\ast}_6$ is the Hodge dual computed with respect to the flat metric $\delta_{\Lambda\Sigma}$ on the transverse $\mb{R}^6$. 

We will demonstrate that \eqref{N=4ansatz} solves the Type IIB Bianchi identities and equations of motion, which are presented in Appendix \ref{IIBeoms}, provided that the four-dimensional fields $(g_{\mu\nu},\phi,\chi,\mc{F}^\Lambda_{(2)})$ solve the Bianchi identities and equations of motion of $\mc{N}=4$ supergravity \eqref{4dbianchi}-\eqref{4dEinstein}. Let's begin with the Bianchi identities. The five-form in \eqref{N=4ansatz} is closed since $H$ is harmonic, so its Bianchi identity \eqref{F5} becomes
\begin{equation}
\hat F_{(3)}\wedge \hat H_{(3)} = 0 \,,
\end{equation}
which is satisfied by our three-form fluxes, as $\mc{F}^\Lambda_{(2)}\wedge{\ast}_4\mc{F}^\Sigma_{(2)}$ is symmetric in $(\Lambda,\Sigma)$. Next, the Bianchi identities (\ref{H3},\ref{F3}) for the three-forms give
\begin{align}\label{d3n=4bianchi}
\hat H_{(3)}:\quad d\mc{F}^\Lambda_{(2)} = 0 \,,\\
\hat F_{(3)}:\quad d\mc{G}^\Lambda_{(2)} = 0 \,,
\end{align}
for all $\Lambda\in\{1,\dots,6\}$, where $\mc{G}^\Lambda_{(2)} = e^{-\phi}{\ast}_4\mc{F}^\Lambda_{(2)} + \chi\mc{F}^\Lambda_{(2)}$ is the $SL(2,\mb{R})$ dual of $\mc{F}^\Lambda_{(2)}$ as defined in \eqref{sl2dual4d}. These are the Bianchi identities and flux equations of motion of the $\mc{N}=4$ theory.

To evaluate the equations of motion, we record below the Hodge duals
\begin{equation}
\begin{split}
&\hat{\ast}d\hat\Phi = H{\ast}_4d\phi\wedge \vol_6 \,,\quad \hat{\ast}\hat F_{(1)} = -H{\ast}_4d\chi\wedge\vol_6 \,,\\
&\hat{\ast}\hat H_{(3)} = \frac{1}{\sqrt{2}}H{\ast}_4\mc{F}^\Lambda_{(2)}\wedge{\ast}_6dy^\Lambda \,,\quad \hat{\ast}\hat F_{(3)} = \frac{1}{\sqrt{2}}He^{-\phi}\mc{F}^\Lambda_{(2)}\wedge{\ast}_6dy^\Lambda \,,
\end{split}
\end{equation}
where $\vol_6$ is the volume form on $\mb{R}^6$. Beginning with the $\hat F_{(3)}$ equation \eqref{eomF3}, we find
\begin{equation}
d(e^{\hat\Phi}{\ast}\hat F_{(3)}) = \frac{1}{\sqrt{2}}\partial_\Lambda H\mc{F}^\Lambda_{(2)}\wedge\vol_6 \,,
\end{equation}
where we have used the Bianchi identity $d\mc{F}^\Lambda_{(2)}=0$ and the identity $d{\ast}_6dy^\Lambda = 0$. Next,
\begin{equation}
\hat F_{(5)} \wedge \hat H_{(3)} = \frac{1}{\sqrt{2}}\partial_\Lambda H\mc{F}^\Lambda_{(2)}\wedge\vol_6 \,,
\end{equation}
so
\begin{equation}
d(e^{\hat\Phi}{\ast}\hat F_{(3)}) - \hat F_{(5)}\wedge \hat H_{(3)} = 0 
\end{equation}
identically. For the $\hat H_{(3)}$ equation \eqref{eomH3}, a similar calculation shows that it reduces to a $0=0$ identity using the Bianchi identities and field equations for $\mc{F}^\Lambda_{(2)}$. For the RR scalar equation, we have
\begin{equation}
d(e^{2\hat\Phi}\hat{\ast}\hat F_{(1)}) = -d(He^{2\phi}{\ast}_4d\chi\wedge\vol_6) = -Hd(e^{2\phi}{\ast}_4d\chi)\wedge\vol_6
\end{equation}
as $dH\wedge\vol_6 = 0$ and $d\vol_6 = 0$, and 
\begin{equation}
\hat H_{(3)}\wedge\hat{\ast}\hat F_{(3)} = \frac{1}{2}He^{-\phi}\mc{F}^\Lambda\wedge\mc{F}^\Sigma dy^\Lambda\wedge{\ast}_6dy^\Sigma = \frac{1}{2}He^{-\phi}\mc{F}^\Lambda_{(2)}\wedge\mc{F}^\Lambda_{(2)}\wedge \vol_6 \,.
\end{equation}
Therefore, \eqref{eomF1} becomes
\begin{equation}
d(e^{2\phi}{\ast}_4d\chi) - \frac{1}{2}\mc{F}^\Lambda_{(2)}\wedge\mc{F}^\Lambda_{(2)} = 0 \,,
\end{equation} 
and we find that the transverse dependence factors out nicely. Similarly, the dilaton equation \eqref{eomPhi} reduces to
\begin{equation}
d{\ast}_4d\phi - e^{2\phi}d\chi\wedge{\ast}_4d\chi + \frac{1}{2}e^{-\phi}\mc{F}^\Lambda_{(2)}\wedge{\ast}_4\mc{F}^\Lambda_{(2)} = 0 \,.
\end{equation}
These are the scalar equations for the $\mc{N}=4$ theory. 

Finally, we move to the Einstein equations \eqref{IIBEinstein}. The non-zero components of the Ricci tensor are
\begin{equation}
\hat R_{\mu\nu} = R_{\mu\nu} -\frac{64\pi^2 N^2}{r^{10}H^3}g_{\mu\nu} \,,\quad \hat R_{\Lambda\Sigma} = -\frac{1}{2}H^{-2}\partial_\Lambda H \partial_\Sigma H+ \frac{64\pi^2 N^2}{r^{10}H^2}\delta_{\Lambda\Sigma} \,.
\end{equation} 
Let 
\begin{equation}
\hat T^1_{MN} = \frac{1}{96}\hat F^{}_{MP_1P_2P_3P_4}\hat F_N^{\ph{M}P_1P_2P_3P_4} \,,
\end{equation}
be the contribution of $\hat F_{(5)}$ to the stress tensor. A straightforward calculation shows that the non-zero components of $\hat T^1_{MN}$ are
\begin{equation}
\hat T^1_{\mu\nu} = -\frac{64\pi^2 N^2}{r^{10}H^3}g_{\mu\nu} \,,\quad \hat T^1_{\Lambda\Sigma} = -\frac{1}{2}H^{-2}\partial_\Lambda H \partial_\Sigma H+ \frac{64\pi^2 N^2}{r^{10}H^2}\delta_{\Lambda\Sigma} \,,
\end{equation}
which perfectly cancels the \say{brane part} of the Ricci tensor. Thus, the Einstein equations reduce to
\begin{equation}\label{worldvolumeEinstein}
\begin{split}
R_{\mu\nu} &= \frac{1}{2} \hat \nabla_\mu \hat\Phi \hat \nabla_\nu \hat\Phi+\frac{1}{2} e^{2\hat\Phi} \hat \nabla_\mu \hat C_{0} \hat\nabla_\nu \hat C_{0} \\
&\quad + \frac{1}{4}e^{-\hat\Phi}\left(\hat H^{}_{\mu P_1P_2} \hat H_{\nu}^{\ph{\nu}P_1P_2}- \frac{1}{12}(\hat H_{(3)})^2\hat g_{\mu\nu} \right)\\
&\quad + \frac{1}{4}e^{\hat\Phi}\left(\hat F^{}_{\mu P_1P_2} \hat F_{\nu}^{\ph{\nu}P_1P_2} - \frac{1}{12}(\hat F_{(3)})^2\hat  g_{\mu\nu}\right) 
\end{split}
\end{equation}
for the worldvolume components,
\begin{equation}\label{mixedEinstein}
\begin{split}
&\frac{1}{2} \hat\nabla_\mu \hat\Phi \hat\nabla_\Sigma \hat\Phi+\frac{1}{2} e^{2\hat\Phi} \hat\nabla_\mu \hat C_{0} \hat \nabla_\Sigma \hat C_{0} \\
&+ \frac{1}{4}e^{-\hat\Phi}\left(\hat H^{}_{\mu P_1P_2} \hat H_{\Sigma}^{\ph{\Sigma}P_1P_2}- \frac{1}{12}(\hat H_{(3)})^2\hat g_{\mu\Sigma} \right)\\
&+ \frac{1}{4}e^{\hat\Phi}\left(\hat F^{}_{\mu P_1P_2} \hat F_{\Sigma}^{\ph{\Sigma}P_1P_2} - \frac{1}{12}(\hat F_{(3)})^2\hat g_{\mu\Sigma}\right) = 0 
\end{split}
\end{equation}
for the mixed components, and
\begin{equation}\label{transverseEinstein}
\begin{split}
&\frac{1}{2} \hat\nabla_\Lambda \hat\Phi \hat\nabla_\Sigma \hat\Phi+\frac{1}{2} e^{2\hat\Phi} \hat\nabla_\Lambda\hat C_{0} \hat\nabla_\Sigma\hat C_{0} \\
&+ \frac{1}{4}e^{-\hat\Phi}\left(\hat H^{}_{\Lambda P_1P_2} \hat H_{\Sigma}^{\ph{\Sigma}P_1P_2}- \frac{1}{12}(\hat H_{(3)})^2\hat g_{\Lambda\Sigma} \right)\\
&+ \frac{1}{4}e^{\hat\Phi}\left(\hat F^{}_{\Lambda P_1P_2} \hat F_{\Sigma}^{\ph{\Sigma}P_1P_2} - \frac{1}{12}(\hat F_{(3)})^2\hat g_{\Lambda\Sigma}\right) = 0 
\end{split}
\end{equation}
for the transverse components. Using the ingredients provided in Appendix \ref{usefulD3}, we find that \eqref{worldvolumeEinstein} reduces to
\begin{equation}
R_{\mu\nu} = \frac{1}{2}\nabla_\mu\phi\nabla_\mu\phi + \frac{1}{2}e^{2\phi}\nabla_\mu\chi\nabla_\nu\chi + \frac{1}{2}e^{-\phi}\left(\mc{F}^\Lambda_{\mu\rho}\mc{F}^{\Lambda\,\rho}_\nu - \frac{1}{4}(\mc{F}^\Lambda)^2g_{\mu\nu}\right) \,,
\end{equation}
This the Einstein equation of the $\mc{N}=4$ theory. The $(\mu,\Lambda)$ and $(\Lambda,\Sigma)$ components of the Einstein equations reduce to $0=0$ identities. 

We have now shown explicitly that \eqref{N=4ansatz} is a consistent truncation of Type IIB supergravity on a background of $N$ parallel D3-branes to the $\mc{N}=4$ supergravity theory on the four-dimensional worldvolume.

\subsubsection{Asymptotic limits}\label{asymptoticlimits}

The skeleton D3-brane solution \eqref{D3branesol} is part of a large class of interpolating solitons. At small $r$, the solution becomes $\hbox{AdS}_5\times S^5$, and at large $r$, it becomes the $\mb{R}^{1,9}$ vacuum. We will examine our embedding of the $\mc{N}=4$ theory obtained in \eqref{N=4ansatz} at these limits, and find that \eqref{N=4ansatz} retains the interpolating structure.

\subsubsubsection{The flat limit}

As $r\to\infty$, the harmonic function $H\to1$, and the embedding of the $\mc{N}=4$ theory \eqref{N=4ansatz} becomes
\begin{equation}\label{flatansatz}
\begin{split}
&d\hat s^2 = g_{\mu\nu}(x)dx^\mu dx^\nu + d{y}^\Lambda d{y}^\Lambda \,,\quad \hat\Phi= \phi(x)\,,\quad \hat C_0 = -\chi(x) \,,\\
&\hat H_{(3)} = \frac{1}{\sqrt{2}}\mc{F}^\Lambda_{(2)}\wedge d{y}^\Lambda \,,\quad \hat F_{(3)} = -\frac{1}{\sqrt{2}}e^{-\phi}{\ast}_4\mc{F}^\Lambda_{(2)}\wedge d{y}^\Lambda \,,\quad \hat F_{(5)} = 0 \,,
\end{split}
\end{equation}
which describes the embedding of the $\mc{N}=4$ theory into the $\mb{R}^{1,9}$ vacuum of Type IIB supergravity. It is easy to check that \eqref{flatansatz} is indeed a solution to the Type IIB equations of motion provided that the four-dimensional fields satisfy the equations of motion of the $\mc{N}=4$ theory.

As an aside, it is interesting to consider compactifying the $\mb{R}^6$ in \eqref{flatansatz} by a compact quotient. This is now possible because the $\mb{R}^6$ flat-space trivialisation of the harmonic function supporting the original solution allows $\partial/\partial y^\Lambda$ to be Killing vectors. The simplest compactification will be the unit six-torus $T^6$, where the coordinates ${y}^\Lambda$ take values in $[0,2\pi)$. With this identification, the Page charges associated to $\hat H_{(3)}$ and $\hat F_{(3)}$, corresponding to the charges of the NS5- and D5-branes sourced by these fluxes, are now finite, with
\begin{equation}
Q_{\text{NS5}} = \frac{1}{4\pi^2}\int_{C_3}\hat H_{(3)} = \frac{1}{\sqrt{2}}\sum_{\Lambda=1}^6\frac{1}{2\pi} \int_{C_2}\mc{F}^\Lambda_{(2)} \,,
\end{equation}
and
\begin{equation}
Q_{\text{D5}} = \frac{1}{4\pi^2}\int_{C'_3}\left(\hat F_{(3)} + \hat C_0\hat H_{(3)}\right) = -\frac{1}{\sqrt{2}}\sum_{\Lambda=1}^6\frac{1}{2\pi}\int_{C'_2}\mc{G}^\Lambda_{(2)} \,,
\end{equation}
where $C_2$ and $C_2'$ are, in principle, different two-cycles. We can write this more succinctly as
\begin{equation}
Q_{\text{NS5}} = \frac{1}{\sqrt{2}}\sum_{\Lambda=1}^6c_1(\mc{F}^\Lambda_{(2)}) \,,\quad Q_{\text{D5}} = -\frac{1}{\sqrt{2}}\sum_{\Lambda=1}^6c_1(\mc{G}^\Lambda_{(2)})\,.
\end{equation}
where $c_1$ denotes the first Chern class. The first Chern classes are quantised, so
\begin{equation}
Q_{\text{NS5}}, Q_{\text{D5}} \in \frac{1}{\sqrt{2}}\mb{Z} \,.
\end{equation}
In string theory, however, we must have $Q_{\text{NS5}}, Q_{\text{D5}}\in\mb{Z}$. To remedy this, we employ a trombone rescaling of the Type IIB fields (see \eqref{tromboneIIB}), 
\begin{equation}
\hat g \mapsto \sqrt{2}\hat g\,,\quad \hat\Phi\mapsto \hat\Phi\,,\quad \hat C_0\mapsto \hat C_0\,,\quad \hat H_{(3)} \mapsto \sqrt{2}\hat H_{(3)} \,,\quad \hat F_{(3)} \mapsto \sqrt{2}\hat F_{(3)} \,,
\end{equation}
so that 
\begin{equation}
Q_{\text{NS5}} = \sum_{\Lambda=1}^6c_1(\mc{F}^\Lambda_{(2)}) \,,\quad Q_{\text{D5}} = -\sum_{\Lambda=1}^6c_1(\mc{G}^\Lambda_{(2)})
\end{equation}
are integers. We note that the NS5 charge corresponds to the magnetic charges of the $\mc{N}=4$ theory, whereas the D5 charge corresponds to the generalised electric charges.

\subsubsubsection{The horizon limit and $\mc{N}=2$ supergravity} 

As $r\to0$, the harmonic function $H\to c/r^4$ with $c=4\pi N$, and the embedding of the $\mc{N}=4$ theory \eqref{N=4ansatz} becomes
\begin{equation}\label{horizonansatz}
\begin{split}
&d\hat s^2 = c^{-1/2}r^2g_{\mu\nu}(x)dx^\mu dx^\nu + c^{1/2}\frac{dr^2}{r^2} + c^{1/2}d\Omega^2_5 \,,\quad \hat\Phi= \phi(x)\,,\quad \hat C_0 = -\chi(x) \,,\\
&\hat H_{(3)} = \frac{1}{\sqrt{2}}\mc{F}^\Lambda_{(2)}\wedge d{y}^\Lambda \,,\quad \hat F_{(3)} = -\frac{1}{\sqrt{2}}e^{-\phi}{\ast}_4\mc{F}^\Lambda_{(2)}\wedge d{y}^\Lambda \,, \\
&\hat F_{(5)} = 16\pi N\left(\vol_5 - c^{-2}r^3\vol_4\wedge dr\right) \,,
\end{split}
\end{equation}
where we have used spherical polar coordinates on $\mb{R}^6$, and we recall that $d\Omega^2_5$ and $\vol_5$ are respectively the metric and the volume form on the round five-sphere. For an asymptotically flat metric $g_{\mu\nu}$, \eqref{horizonansatz} corresponds to a truncation of Type IIB supergravity on the five-sphere down to an asymptotically $\hbox{AdS}_5$ spacetime. It can be verified that \eqref{horizonansatz} is also an exact solution to the Type IIB equations of motion provided that the four-dimensional fields obey the corresponding equations of motion of the $\mc{N}=4$ theory.

The $\mc{N}=4$ supergravity theory \eqref{goaltheory} in four dimensions admits a consistent truncation to the $\mc{N}=2$ supergravity theory, which contains one graviton, one vector, and two gravitini. The Lagrangian for the bosonic sector of this theory is simply the Einstein-Maxwell Lagrangian,
\begin{equation}\label{N=2theory}
\mc{L}_4 = R{\ast}_41 - \frac{1}{2}F_{(2)}\wedge{\ast}_4F_{(2)}\,.
\end{equation}
There are multiple ways of getting to the $\mc{N}=2$ theory from the $\mc{N}=4$ theory, but at least two of the vectors of the $\mc{N}=4$ theory must be retained. Keeping only $\mc{F}_{(2)}^5$ and $\mc{F}_{(2)}^6$, the truncation is given by
\begin{equation}\label{N=2truncationfromN=4}
\phi = \chi = 0 \,,\quad \mc{F}_{(2)}^5 = \frac{1}{\sqrt{2}}{\ast}_4F^{}_{(2)} \,,\quad \mc{F}_{(2)}^{6} = \frac{1}{\sqrt{2}}F^{}_{(2)} \,.
\end{equation}
In terms of the Type IIB embedding \eqref{horizonansatz}, we have
\begin{equation}\label{horizonansatzN=2}
\begin{split}
&d\hat s^2 = c^{-1/2}r^2g_{\mu\nu}(x)dx^\mu dx^\nu + c^{1/2}\frac{dr^2}{r^2} + c^{1/2}d\Omega^2_5 \,,\quad \hat\Phi= 0\,,\quad \hat C_0 = 0\,,\\
&\hat H_{(3)} = \frac{1}{2}\left(F_{(2)}\wedge dy^6 + {\ast}_4F_{(2)}\wedge dy^5\right)\,,\quad \hat F_{(3)} = \frac{1}{2}\left(F_{(2)}\wedge dy^5 - {\ast}_4F_{(2)}\wedge dy^6\right) \,,\\
&\hat F_{(5)} = 16\pi N\left(\vol_5 - c^{-2}r^3\vol_4\wedge dr\right) \,.
\end{split}
\end{equation}
The $\mc{N}=2$ theory was also embedded into the $\hbox{AdS}_5\times S^5$ near-horizon of parallel D3-branes in Ref.\cite{Lu:2000xc} using a specific system of spherical polar coordinates. We shall now show that our embedding \eqref{horizonansatzN=2} is equivalent to that of Ref.\cite{Lu:2000xc}. 

In the absence of the Type IIB scalars, it is convenient to combine the NSNS and RR three-forms into a complex three-form $\hat G_{(3)} = \hat F_{(3)}+i\hat H_{(3)}$. The Bianchi identities and equations of motion of the Type IIB theory in terms of $\hat G_{(3)}$ are given in Appendix \ref{IIBeoms}. Choosing the following spherical coordinates on $\mb{R}^6$,
\begin{equation}
y^{1,2,3,4} = rv^{1,2,3,4}\cos\xi  \,,\quad y^5 = r\sin\xi\cos\tau\,,\quad y^6 = r\sin\xi\sin\tau\,,
\end{equation}
where $v^i$, $i\in\{1,2,3,4\}$, parametrise the unit three-sphere, satisfying $v^iv^i=1$. Then, we find that $\hat G_{(3)}$ is locally exact, $\hat G_{(3)} = d\hat A_{(2)}$, with 
\begin{equation}
\hat A_{(2)} = \frac{1}{2}e^{i\tau}r\sin\xi\left(F_{(2)} + i{\ast}_4F_{(2)}\right) \,.
\end{equation}
Performing the change of coordinates $r = e^{-k|z|}$, we recover the embedding of Ref.\cite{Lu:2000xc}. Consequently, the construction of Ref.\cite{Lu:2000xc} represents an $\mc N=2$ truncated asymptotic limit as $r\to0$ of the full D3-brane $\mc N=4$ embedding of the present paper.

\subsection{Stationary solutions and black holes}\label{BHsolutions}

The $\mc{N}=4$ supergravity theory contains many supersymmetric, stationary black hole solutions. Using our embedding \eqref{N=4ansatz}, these black holes can be uplifted to Type IIB supergravity, where they then have the interpretation of being situated on worldvolumes of D3-branes. These black holes can be constructed by first reducing the four-dimensional theory along a timelike $U(1)$ isometry to a Euclidean theory in three dimensions, whose bosonic sector consists of a metric coupled to non-linear sigma model with a pseudo-Riemannian target space $\mc{M}_S$. Following the work of \cite{Neugebauer:1969wr,Breitenlohner:1987dg,Clement:1996nh,Galtsov:1998mhf}, null geodesics on $\mc{M}_S$ correspond to certain classes of black holes. Here, we will perform the timelike Kaluza-Klein reduction explicitly and show that the corresponding target space is $\mc{M}_S = SO(8,2)/SO(6,2)\times SO(2)$. A class of null geodesics will also be constructed on a submanifold $\mc{N}_S = \mb{R}\times SO(2,1)/SO(1,1)\subset\mc{M}_S$ for completeness.

Consider a timelike reduction of four-dimensional $\mc{N}=4$ supergravity,
\begin{equation}\label{trunc3d}
\begin{split}
&ds^2_4 = -e^{\rho(x)}(dt + A_{(1)})^2 + e^{-\rho(x)}h_{mn}(x)dx^mdx^n \,,\quad \phi=\phi(x)\,,\quad \chi = \chi(x) \,,\\
&\mc{A}_{(1)}^\Lambda = a^\Lambda(x)(dt + A_{(1)}) + \mc{B}_{(1)}^\Lambda \,,
\end{split}
\end{equation}
where $m,n\in\{1,2,3\}$. Defining $\mc{H}_{(2)}^\Lambda = d\mc{B}^\Lambda_{(1)} + a^\Lambda F_{(2)}$, where $F_{(2)}= dA_{(1)}$, the equations of motion can be obtained from the Lagrangian
\begin{equation}
\begin{split}
\mc{L}_3 &= R{\ast}_31-\frac{1}{2}d\phi\wedge{\ast}_3d\phi - \frac{1}{2}e^{2\phi}d\chi\wedge{\ast}_3d\chi -\frac{1}{2}d\rho\wedge{\ast}_3d\rho + \frac{1}{2}e^{-\phi-\rho}da^\Lambda\wedge{\ast}_3da^\Lambda \\
&\quad + \frac{1}{2}e^{2\rho}F_{(2)}\wedge{\ast}_3F_{(2)} - \frac{1}{2}e^{-\phi+\rho}\mc{H}_{(2)}^\Lambda\wedge{\ast}_3\mc{H}^\Lambda_{(2)} + \chi\,da^\Lambda\wedge\mc{H}_{(2)}^\Lambda \,,
\end{split}
\end{equation}
where ${\ast}_3$ is the Hodge dual defined with respect to the three-dimensional metric $h_{mn}$, and $R$ is the Ricci scalar of $h_{mn}$. In three dimensions, vectors are dual to scalars, so the spectrum of this Euclidean theory consists of the metric $h_{mn}$ and sixteen scalars. To construct this vector-scalar dualisation, we first examine the Bianchi identities and the equations of motion of $\mc{H}^\Lambda_{(2)}$. They are
\begin{align}
&d\mc{H}_{(2)}^\Lambda = da^\Lambda\wedge F_{(2)}\,,\label{3dHbianchi} \\
&d(e^{-\phi+\rho}{\ast}_3\mc{H}_{(2)}^\Lambda - \chi\,da^\Lambda)  = 0 \,. \label{3dHeqn}
\end{align}
The dual formulation is in terms of six scalars $h^\Lambda$ defined by
\begin{equation}\label{3dHdual}
\mc{H}_{(2)}^\Lambda = e^{\phi-\rho}{\ast}_3(dh^\Lambda + \chi\,da^\Lambda) \,,
\end{equation}
so that \eqref{3dHeqn} is trivially satisfied. \eqref{3dHbianchi} then becomes the equation of motion for $h^\Lambda$. Next, we examine the Bianchi identity and the equation of motion of $F_{(2)}$,
\begin{align}
&dF_{(2)} = 0 \,,\label{3dFbianchi} \\
&d(e^{2\rho}{\ast}_3F_{(2)}) - da^\Lambda\wedge dh^\Lambda = 0 \,,
\end{align}
where we have used \eqref{3dHdual} to write $\mc{H}^\Lambda_{(2)}$ in terms of $h^\Lambda$. These can be dualised by introducing a scalar $f$, with
\begin{equation}
F_{(2)} = -e^{-2\rho}{\ast}_3\left(df + \frac{1}{2}h^\Lambda da^\Lambda-\frac{1}{2}a^\Lambda dh^\Lambda\right) \,.
\end{equation}
In the $f$ and $h^\Lambda$ variables, the Lagrangian reads
\begin{equation}
\begin{split}
\mc{L}_3 &= R{\ast}_31-\frac{1}{2}d\phi\wedge{\ast}_3d\phi - \frac{1}{2}e^{2\phi}d\chi\wedge{\ast}_3d\chi -\frac{1}{2}d\rho\wedge{\ast}_3d\rho  \\
&\quad + \frac{1}{2}e^{-\phi-\rho}da^\Lambda\wedge{\ast}_3da^\Lambda+ \frac{1}{2}e^{\phi-\rho}\mc{P}_{(1)}^\Lambda\wedge{\ast}_3\mc{P}_{(1)}^\Lambda - \frac{1}{2}e^{-2\rho}\mc{Q}_{(1)}\wedge{\ast}_3\mc{Q}_{(1)}\,,
\end{split}
\end{equation}
where we have defined the one-forms
\begin{equation}
\mc{P}_{(1)}^\Lambda = dh^\Lambda + \chi\,da^\Lambda\,,\quad \mc{Q}_{(1)} = df + \frac{1}{2}h^\Lambda da^\Lambda-\frac{1}{2}a^\Lambda dh^\Lambda \,.
\end{equation}
As discussed in \cite{Hull:1998br,Lu:1998xt}, the scalars parametrise the coset space $SO(8,2)/\left(SO(6,2)\times SO(2)\right)$ with signature $(\underbrace{+,\dots,+}_{4},\underbrace{-,\dots,-}_{12})$, which we will now show explicitly. Firstly, we will perform the field redefinition
\begin{equation}
\rho = -\frac{1}{\sqrt{2}}(\phi_1+\phi_2)\,,\quad \phi =-\frac{1}{\sqrt{2}}(\phi_1-\phi_2) \,,
\end{equation}
so that the Lagrangian reads
\begin{equation}\label{3dLag}
\begin{split}
\mc{L}_3 &= R{\ast}_31-\frac{1}{2}d\phi_1\wedge{\ast}_3d\phi_1 - \frac{1}{2}d\phi_2\wedge{\ast}_3d\phi_2 - \frac{1}{2}e^{-\sqrt{2}\phi_1+\sqrt{2}\phi_2}d\chi\wedge{\ast}_3d\chi \\
&\quad + \frac{1}{2}e^{\sqrt{2}\phi_1}da^\Lambda\wedge{\ast}_3da^\Lambda+ \frac{1}{2}e^{\sqrt{2}\phi_2}\mc{P}_{(1)}^\Lambda\wedge{\ast}_3\mc{P}_{(1)}^\Lambda - \frac{1}{2}e^{\sqrt{2}\phi_1+\sqrt{2}\phi_2}\mc{Q}_{(1)}\wedge{\ast}_3\mc{Q}_{(1)}\,.
\end{split}
\end{equation}
Then, using the results of \cite{Lu:1998xt}, we find that \eqref{3dLag} can be written as
\begin{equation}\label{3dLagCoset}
\mc{L}_3 = R{\ast}_31 + \frac{1}{4}\tr (dM^{-1}\wedge{\ast}_3dM)\,,
\end{equation}
with $M = \mc{V}^TW_0\mc{V}$, where
\begin{equation}
\mc{V} = \exp\left(\frac{1}{2}\phi_1H_1+\frac{1}{2}\phi_2H_2\right)\exp\left(-\chi E_1^{\ph{1}2}\right)\exp\left(-fV^{12}\right)\exp\left(a^\Lambda U^{\ph{\Lambda}1}_\Lambda+h^\Lambda U^{\ph{\Lambda}2}_\Lambda\right)
\end{equation}
is the coset representative,\footnote{We identify our field variables with those of \cite{Lu:1998xt} by $\mc{A}^1_{(0)2} = -\chi, A_{(0)12} = -f, B^\Lambda_{(0)1} = a^\Lambda, B^\Lambda_{(0)2} = h^\Lambda$, and replace their $I$ index with our $\Lambda$ index.} $(H_1,H_2)$ and $(E_1^{\ph{1}2},V^{12},U^{\ph{\Lambda}1}_\Lambda,U^{\ph{\Lambda}2}_\Lambda)$ are respectively the two non-compact Cartan generators and the fourteen positive-root generators of $\mathfrak{so}(2,8)$ in the notation of \cite{Lu:1998xt}, and the fiducial matrix $W_0$ is defined as
\begin{equation}
W_0 = \diag(-1,-1,1,1,1,1,1,1,-1,-1) \,.
\end{equation}
The structure of $W_0$ determines the denominator group of the coset to be $SO(6,2)\times SO(2)$. 

We have thus shown that all stationary solutions to the $\mc{N}=4$ supergravity theory are encoded in a three-dimensional Euclidean theory describing gravity minimally coupled to sixteen scalars parametrising the coset space $SO(8,2)/\left(SO(6,2)\times SO(2)\right)$. A special class of these stationary solutions are black hole solutions which correspond to harmonic maps between three-dimensional Euclidean space and the coset space \cite{Neugebauer:1969wr,Breitenlohner:1987dg,Clement:1996nh,Galtsov:1998mhf}. From \eqref{3dLagCoset}, the equations of motion for the metric $h_{mn}$ and the scalars $M$ are 
\begin{equation}
\begin{split}
&R_{mn} = -\frac{1}{4}\tr\left(\nabla_mM\nabla_nM^{-1}\right) \,,\\
&\nabla_m\left(M^{-1}\nabla^m M \right) = 0 \,,
\end{split}
\end{equation}
where $\nabla$ is the Levi-Civita connection with respect to $h_{mn}$. A key observation of \cite{Neugebauer:1969wr} was that these equations simplify greatly if the scalars are taken to depend on the three spatial coordinates through a single, harmonic function $\sigma(x)$. In this case, the scalar equations become
\begin{equation}
\frac{d}{d\sigma}\left(M^{-1}\frac{dM}{d\sigma}\right) = 0\,,
\end{equation}
which is a geodesic equation on the scalar manifold. The solution to this is
\begin{equation}
M = A\exp(\sigma B) \,,
\end{equation}
where $A\in SO(8,2)/SO(6,2)\times SO(2)$, and $B \in \mathfrak{so}(8,2)$ are constant matrices. Substituting this into the Einstein equations gives
\begin{equation}
R_{mn} = \frac{1}{4}\nabla_m\sigma\nabla_n\sigma\tr(B^2) \,,
\end{equation}
and we note that the metric on the scalar target manifold along the geodesic becomes
\begin{equation}
ds^2 = -\frac{1}{2}\tr (dM^{-1}dM) = \frac{1}{2}\tr(B^2)d\sigma^2 \,.
\end{equation}
Thus, scalar-manifold null geodesics (which exist due to the split signature on it) are mapped-to from three-dimensional Ricci-flat geometries. When lifted back to ten dimensions using \eqref{N=4ansatz}, such solutions correspond to D3-branes with charged black holes on their worldvolume. In three dimensions, baring conical singularities, Ricci-flat geometries are flat, so null geodesics on the scalar target manifold are mapped-to from $\mb{R}^3$. The harmonic function $\sigma(x)$ takes the form
\begin{equation}
\sigma(x) =\sum_{l} \frac{k^l}{|x - x_l|} \,,
\end{equation} 
where $k^l$ are constants, and we have imposed the boundary condition $\sigma(\infty) = 0$ so that the solutions when lifted back to four dimensions are asymptotically flat. This construction can be generalised to include multiple harmonic functions $\sigma_a$.

For illustration, we will now construct a simple null geodesic on the scalar target manifold. It is consistent to keep only $\phi_1,\phi_2$ and one of the $a^\Lambda$, say $a^1 = a$. The metric on the scalar manifold then becomes
\begin{equation}
ds^2 = d\phi_1^2 + d\phi_2^2 - e^{\sqrt{2}\phi_1}da^2 \,,
\end{equation}
which is the metric on  $\mb{R}\times SO(2,1)/SO(1,1)$. The geodesic equations are then
\begin{equation}
\begin{split}
\phi_1:\quad &\phi_1'' + \frac{1}{\sqrt{2}}e^{\sqrt{2}\phi_1}(a')^2 = 0\,,\\
\phi_2:\quad &\phi_2'' = 0 \,,\\
a:\quad &a'' + \sqrt{2}a'\phi_1' = 0 \,,
\end{split}
\end{equation}
where the primes denote derivatives with respect to the harmonic function $\sigma$. The general solution for the $\phi_2$ equation is
\begin{equation}
\phi_2(\sigma) = p_0 + p_1\sigma\,,
\end{equation}
where $p_{0,1}$ are constants. With this, the null condition reads
\begin{equation}\label{nullcondition}
(a')^2 = e^{-\sqrt{2}\phi_1}\left((\phi_1')^2 + p_1^2\right) \,,
\end{equation}
which when substituted into the $\phi_1$ (or $a$) equation, gives
\begin{equation}
\phi_1'' + \frac{1}{\sqrt{2}}(\phi_1')^2 + \frac{p_1^2}{\sqrt{2}} = 0 \,.
\end{equation}
Performing the field redefinition $u(\sigma) = e^{\phi_1(\sigma)/\sqrt{2}}$, this becomes the eigenvalue equation
\begin{equation}
u'' = -\frac{p_1^2}{2}u \,,
\end{equation}
the general solution of which is
\begin{equation}
u(\sigma) = q_0\cos\left(\frac{p_1}{\sqrt{2}}\sigma\right) + \frac{q_1}{p_1}\sin\left(\frac{p_1}{\sqrt{2}}\sigma\right)\,,
\end{equation}
where $q_{0,1}$ are constants. The value of $a(\sigma)$ can then be obtained through \eqref{nullcondition}. In the limit $p_1\to0$, which corresponds to consistently truncating out the dynamics of $\phi_2$, we have
\begin{equation}
\lim_{p_1\to0}u(\sigma) = q_0 + \frac{q_1}{\sqrt{2}}\sigma \,.
\end{equation}

\section{M5-branes with their worldvolume supergravities}\label{sec:M5}

The geometry of the $\frac12$-BPS solution of M-theory describing parallel M5-branes, or in the present language, skeleton M5-branes, is a warped product $\mb{R}^{1,5}\times_W\mb{R}^5$. As with the skeleton D3-branes of the previous section, it is perfectly consistent to replace the $\mb{R}^{1,5}$ worldvolume with a generic, six-dimensional manifold $M_6$ with a Ricci-flat metric $g_{\mu\nu}$. Since the original solution preserves sixteen supercharges and since the Killing spinor is chiral on the six-dimensional worldvolume, we may now anticipate that there is in fact a consistent truncation to pure worldvolume $\mc{N}=(4,0)$ supergravity in six dimensions. In this section, we will construct the corresponding embedding of six-dimensional, $\mc{N}=(4,0)$ supergravity onto the worldvolume of the skeleton M5-branes. We will then consider anti-self-dual string solutions of the $\mc{N}=(4,0)$ theory, and will examine their dynamics when uplifted back into M-theory using the braneworld embedding.

\subsection{Skeleton M5-branes and $d=6$, $\mc{N}=(4,0)$ supergravity}

The M-theory solution describing $N$ parallel M5-branes in units where the M-theory length scale has $l_{11} = 1$ is given by
\begin{equation}\label{pureM5}
d\hat s^2_{11} = H^{-1/3}ds^2(\mb{R}^{1,5}) + H^{2/3}dy^idy^i \,,\quad \hat F_{(4)} = -{\ast}_5dH \,,\quad H = 1+\frac{\pi N}{r^3} \,,
\end{equation}
where $i\in\{1,\dots,5\}$, $r^2 = y^iy^i$, and ${\ast}_5$ is the Hodge dual with respect to the flat metric on the transverse $\mb{R}^5$. For this to be a valid background for a quantum theory, the four-form flux must obey the quantisation condition,
\begin{equation}
\frac{1}{8\pi^3}\int_{C_4}\hat F_{(4)} \in \mb{Z} \,,
\end{equation}
for any non-trivial four-cycle $C_4$. The only non-trivial four-cycle in \eqref{pureM5} is $C_4 = S^4$, the asymptotic four-sphere on the transverse $\mb{R}^5$, and for it we find
\begin{equation}
\frac{1}{8\pi^3}\int_{S^4}\hat F_{(4)} = N \in \mb{Z} \,.
\end{equation}
The solution \eqref{pureM5} is the skeleton on which we will embed the pure, chiral, six-dimensional, $\mc{N}=(4,0)$ supergravity. The bosonic sector of this theory contains a graviton and five two-form potentials with anti-self-dual field strengths. Denoting these bosonic degrees of freedom by $g_{\mu\nu}$ and $B^i_{(2)}$, the equations of motion of the $\mc{N}=(4,0)$ theory are
\begin{eqnarray}
&&R_{\mu\nu} = \frac{1}{4}G^i_{\mu\rho\sigma}G_{\nu}^{i\,\rho\sigma} \,, \label{6deinstein} \\
&&d{\ast}_6G^i_{(3)} = 0 \,,\quad G^i_{(3)} = -{\ast}_6G^i_{(3)} \label{6dG3eom}\,,
\end{eqnarray}
where $i\in\{1,\ldots,5\}$, $G^i_{(3)} = dB^i_{(2)}$, and ${\ast}_6$ is the Hodge dual with respect to the six-dimensional metric $g_{\mu\nu}$.

\subsection{Embedding $\mc{N}=(4,0)$ supergravity on M5-branes}

We find, using similar techniques to the case of the D3-branes, the following embedding of the $\mc{N}=(4,0)$ supergravity theory on the worldvolume of the skeleton M5-branes,
\begin{equation}\label{M5embedding}
\begin{split}
&d\hat s^2_{11} = H^{-1/3}g_{\mu\nu}dx^\mu dx^\nu + H^{2/3}dy^idy^i \,,\\
&\hat F_{(4)} = G^i_{(3)}\wedge dy^i - {\ast}_5dH \,,\quad G^i_{(3)} = -{\ast}_6G^i_{(3)} \,,
\end{split}
\end{equation}
where we impose the anti-self-duality of $G^i_{(3)}$ by hand. Let's begin with the Bianchi identity \eqref{bianchiF4}. We find
\begin{equation}
d\hat F_{(4)} = dG^i_{(3)}\wedge dy^i \,,
\end{equation}
since $d{\ast}_5dH = 0$ by the harmonicity of $H$. The closure of $\hat F_{(4)}$ then requires
\begin{equation}
dG^i_{(3)} = 0 \,,
\end{equation}
for all $i\in\{1,\dots,5\}$. By the anti-self-duality of $G^i_{(3)}$, this is equivalent to its equation of motion \eqref{6dG3eom}. To compute the equation of motion for $\hat F_{(4)}$ \eqref{eomF411d}, we first record the Hodge dual
\begin{equation}
\hat{\ast}\hat F_{(4)} = HG^i_{(3)}\wedge{\ast}_5dy^i + H^{-2}\vol_6\wedge dH \,,
\end{equation}
where $\vol_6$ is the volume form with respect to the six-dimensional metric $g_{\mu\nu}$, and we used the anti-self-duality of $G^i_{(3)}$. Then, using the closure of $G^i_{(3)}$, we find
\begin{equation}
d\hat{\ast}\hat F_{(4)} = -G^i_{(3)}\wedge dH\wedge{\ast}_5dy^i \,.
\end{equation}
The Chern-Simons term reads
\begin{equation}
\hat F_{(4)}\wedge\hat F_{(4)} = -2G^i_{(3)}\wedge dy^i\wedge{\ast}_5dH - G^i_{(3)}\wedge G^j_{(3)}\wedge dy^i\wedge dy^j \,.
\end{equation}
The anti-self-duality of $G^i_{(3)}$ means that $G^i_{(3)}\wedge G^j_{(3)} = -G^i_{(3)}\wedge{\ast}_6G^j_{(3)}$, which is symmetric under exchange of $(i,j)$, so the second term vanishes. This leaves the first term, which using the symmetry of the Hodge dual, can be rewritten as
\begin{equation}
\hat F_{(4)}\wedge\hat F_{(4)} = -2G^i_{(3)}\wedge dH\wedge{\ast}_5dy^i \,.
\end{equation}
Therefore,
\begin{equation}
d\hat{\ast}\hat F_{(4)} -\frac{1}{2}\hat F_{(4)}\wedge\hat F_{(4)}  = 0 
\end{equation}
identically. Unlike the skeleton M5-branes, the Chern-Simons contribution to the flux equation of motion is non-zero, and is crucial for the consistency of our truncation.

We now turn to the Einstein equations \eqref{11einstein}. The non-zero components of the Ricci tensor are
\begin{equation}
\hat R_{\mu\nu} = R_{\mu\nu} - \frac{3\pi^2N^2}{2r^8H^3}g_{\mu\nu}\,,\quad \hat R_{ij} = -\frac{1}{2}H^{-2}\partial_iH\partial_jH + \frac{3\pi^2N^2}{r^8H^2}\delta_{ij} \,.
\end{equation}
where $R_{\mu\nu}$ is the Ricci tensor of $g_{\mu\nu}$. Using the ingredients in Appendix \ref{usefulM5}, we find that the non-zero components of the stress tensor are
\begin{equation}
\hat T_{\mu\nu} = \frac{1}{4}G^i_{\mu\rho\sigma}G_{\nu}^{i\,\rho\sigma} - \frac{3\pi^2N^2}{2r^8H^3}g_{\mu\nu}\,,\quad \hat T_{ij}= -\frac{1}{2}H^{-2}\partial_iH\partial_jH + \frac{3\pi^2N^2}{r^8H^2}\delta_{ij} \,.
\end{equation}
The Einstein equations in the worldvolume directions then reduce to
\begin{equation}
R_{\mu\nu} = \frac{1}{4}G^i_{\mu\rho\sigma}G_{\nu}^{i\,\rho\sigma} \,,
\end{equation}
which is the Einstein equation of the $\mc{N}=(4,0)$ theory \eqref{6deinstein}. The Einstein equations along the $(\mu,i)$ and $(i,j)$ directions reduce to $0=0$ identities. We have thus shown that \eqref{M5embedding} is a consistent truncation to the $\mc{N}=(4,0)$ theory.

\subsubsection{Asymptotic limits}

As is the case for the skeleton D3-branes, skeleton M5-branes are interpolating solitons. At distances far away from the brane, as $|y^i|\to\infty$, the solution \eqref{pureM5} becomes the $\mb{R}^{1,10}$ vacuum, and close to the brane, $|y^i|\to0$, the solution becomes $\hbox{AdS}_7\times S^4$. We will again examine our embedding \eqref{M5embedding} at these limits, and find that \eqref{M5embedding} retains this interpolating structure. 

\subsubsubsection{The flat limit} 

As $|y^i|\to\infty$, the harmonic function $H\to1$, and the embedding \eqref{M5embedding} becomes
\begin{equation}\label{M5embeddingflat}
\begin{split}
&d\hat s^2_{11} = g_{\mu\nu}dx^\mu dx^\nu + dy^idy^i \,,\\
&\hat F_{(4)} = G^i_{(3)}\wedge dy^i \,,\quad G^i_{(3)} = -{\ast}_6G^i_{(3)} \,.
\end{split}
\end{equation}
It can be easily checked that this is a solution to the M-theory equations of motion provided that the six-dimensional fields solve the equations of motion of the $\mc{N}=(4,0)$ theory. Similar to the D3-branes, it is useful to compactify the transverse $\mb{R}^5$, which again is possible because $\partial/\partial y^i$ are Killing vectors due to the trivialisation of the harmonic function. Choosing this compactification to be $T^5$, so $y^i\in[0,2\pi)$ for all $i\in\{1,\dots,5\}$, we find that the quantisation of $\hat F_{(4)}$ becomes
\begin{equation}
\frac{1}{8\pi^3}\int_{C_4}\hat F_{(4)} = \frac{1}{4\pi^2}\sum_{i=1}^5\int_{C_3}G^i_{(3)} \in \mb{Z} \,,
\end{equation}
which is guaranteed as the middle term is just a sum of the quantisation conditions of $G^i_{(3)}$ in the six-dimensional $\mc{N}=(4,0)$ theory. 

\subsubsubsection{The horizon limit and $\mc{N}=(2,0)$ supergravity} 

As $|y^i|\to0$, the harmonic function $H \to c/ r^3$ with $c=\pi N$, and the embedding \eqref{M5embedding} becomes
\begin{equation}\label{M5embeddinghorizon}
\begin{split}
&d\hat s^2_{11} = c^{-1/3}r g_{\mu\nu}dx^\mu dx^\nu + c^{2/3}\frac{dr^2}{r^2} + c^{2/3}\left(d\xi^2+\cos^2\xi\,d\Omega^2_3\right) \,,\\
&\hat F_{(4)} = G^i_{(3)}\wedge dy^i- {\ast}_5dH \,,\quad G^i_{(3)} = -{\ast}_6G^i_{(3)} \,.
\end{split}
\end{equation}
where we have used the following spherical polar coordinates on $\mb{R}^5$,
\begin{equation}
y^{a} = rv^{a}\cos\xi\,,\quad y^5 = r\sin\xi \,,\quad a\in\{1,\dots,4\}\,,
\end{equation}
with $v^a$ parametrising a three-sphere $v^av^a=1$. For an asymptotically flat metric $g_{\mu\nu}$, \eqref{M5embeddinghorizon} describes a truncation of M-theory on a four-sphere to an asymptotically $\hbox{AdS}_7$ spacetime. This is an exact solution of the M-theory equations of motion provided that the six-dimensional fields obey the equations of motion of the $\mc{N}=(4,0)$ theory.

The $\mc{N}=(4,0)$ supergravity theory admits a consistent truncation to minimal $\mc{N}=(2,0)$ supergravity, the bosonic sector of which consists of a metric and a two-form potential with an anti-self-dual flux. The truncation is obtained by setting four of the $G^i_{(3)}$ to zero. Without loss of generality, we will choose $G^5_{(3)} = G^{}_{(3)}$ to be the non-zero one, so \eqref{M5embeddinghorizon} becomes
\begin{equation}\label{M5embeddinghorizontrunc}
\begin{split}
&d\hat s^2_{11} = c^{-1/3}r g_{\mu\nu}dx^\mu dx^\nu + c^{2/3}\frac{dr^2}{r^2} + c^{2/3}\left(d\xi^2+\cos^2\xi\,d\Omega^2_3\right) \,,\\
&\hat F_{(4)} = G_{(3)}\wedge(\sin\xi\,dr+r\cos\xi\,d\xi) +3c\cos^3\xi\,d\xi\wedge\vol_3 \,,\quad G_{(3)} = -{\ast}_6G_{(3)} \,.
\end{split}
\end{equation}
where $\vol_3$ is the volume form on the three-sphere. Recalling that $G_{(3)}$ is closed, we can integrate $\hat F_{(4)}$ to obtain the three-form gauge potential
\begin{equation}
\hat A_{(3)} = r\sin\xi\,G_{(3)} + c\sin\xi(2+\cos^2\xi)\vol_3 \,.
\end{equation}
After performing a coordinate transformation $r = e^{-2k|z|}$, this is precisely the embedding of $\mc{N}=(2,0)$ supergravity on the horizon of M5-branes constructed in \cite{Lu:2000xc}.

\subsection{Anti-self-dual strings}

The $\mc{N}=(4,0)$ supergravity theory admits the multi-charged anti-self-dual string solution
\begin{equation}
ds^2_6 = h^{-1}(-dt^2+dx^2) + h(dR^2 +R^2d\Omega^2_3) \,,\quad G^i_{(3)} = 2q^i(1-{\ast}_6)\vol_3\,,\quad h = 1+\frac{|q|}{R^2} \,,
\end{equation}
where $|q|^2=q^iq^i$, and $\vol_3$ is the volume form on the round three-sphere. The charges are normalised so that
\begin{equation}
\frac{1}{4\pi^2}\int_{S^3} G^i_{(3)} = q^i \in \mb{Z} \,.
\end{equation}
As $R\to0$, the solution asymptotes to $\hbox{AdS}_3\times S^3$, where the three-sphere radius is proportional to $\sqrt{|q|}$, and as $R\to\infty$, the solution asymptotes to $\mb{R}^{1,5}$.

Using our embedding \eqref{M5embedding}, we can construct the anti-self-dual-string-M5 solution, which can be viewed as a configuration of open membranes ending on the skeleton M5-branes. This solution also exhibits an \say{asymptotically asymptotically} $\hbox{AdS}$ structure. There is an eleven-dimensional asymptotic, $|y^i|\to0$, which brings the eleven-dimensional geometry into an $M_7\times S^4$ structure, where $M_7$ is asymptotically $\hbox{AdS}_7$ provided that the six-dimensional metric $g_{\mu\nu}$ is asymptotically flat. There is also a six-dimensional asymptotic, $R\to0$, with which the six-dimensional metric becomes $\hbox{AdS}_3\times S^3$. So, our solution encompasses quite a few different $\hbox{AdS}$ solutions of different dimensions. A particular configuration is to take $|y^i|\to0$ and let the six-dimensional radius $R$ vary. The resulting solution describes a flow between the $\hbox{AdS}_7$ vacuum and a warped $\hbox{AdS}_3$ solution.

\section{Diagonal and vertical reductions}\label{sec:diagvert}

Descendants of the D3- and M5-branes can be obtained by diagonal or vertical dimensional reductions \cite{Duff:1987bx,Lu:1996mg}. Diagonal reductions correspond to Kaluza-Klein reductions in worldvolume directions, and vertical reductions correspond to Kaluza-Klein reductions on the transverse space after creating a transverse-space shift isometry by stacking the branes appropriately. Let's first consider the diagonal reductions. A one-step diagonal reduction of the M5-brane produces a D4-brane, and a one-step diagonal reduction of the D3-brane can be interpreted as a D2-brane via T-duality. One can further apply the diagonal reduction machinery to the D2-brane to obtain a D1-string, again via T-duality. The D1-string can then be dualised to an F1-string using S-duality. We will show in this section that the same logic applies to the embedding of worldvolume supergravities on branes. By substituting the Kaluza-Klein ans\"atze for the worldvolume supergravities into their D3- and M5-brane embeddings, we obtain, through the usual M-theory/IIA duality and IIA/IIB T-duality, the corresponding embeddings of the worldvolume supergravities for the descendant branes.

For vertical reductions, the situation is rather simpler. Mathematically, the consequence of stacking the branes to create an isometry is just to limit the harmonic function so as to depend on one fewer transverse coordinate. For the skeleton branes, this is possible because the only condition on the harmonic function is its harmonicity, which can be satisfied even if the function depends on fewer coordinates. In the examples of the D3- and M5-branes with their worldvolume supergravities embedded, we find that the embeddings \eqref{N=4ansatz} and \eqref{M5embedding} also support vertical reductions in the same way -- the harmonic function just needs to depend on fewer coordinates. This allows us to construct embeddings of worldvolume supergravities on branes that are vertically descendent from the D3- and M5-branes. 

In the following, we will illustrate the diagonal and vertical reductions using the M5-brane with its embedded worldvolume supergravity.

\subsection{Diagonal example: $\text{M5}\to\text{D4}$}

The diagonal descendant of the M5-brane, the D4-brane, has a five-dimensional worldvolume. The worldvolume supergravity theory is then pure $\mc{N}=4$ supergravity in five dimensions, whose bosonic sector consists of a metric $g_{\mu\nu}$, a scalar $\phi$ and six vectors $A^0_{(1)}$, $A^i_{(1)}$ where $i\in\{1,\dots,5\}$. We will now show that this can be obtained by a standard Kaluza-Klein reduction of the M5-brane's worldvolume supergravity, the $\mc{N}=(4,0)$ supergravity theory in six dimensions. Recall that the $\mc{N}=(4,0)$ supergravity theory contains a graviton and five two-form potentials with anti-self-dual fluxes. The Kaluza-Klein ans\"atze is
\begin{equation}
d\hat s^2_6 = \Sigma\,ds^2_5 + \Sigma^{-3}(d\theta + \sqrt{2}{A}^0_{(1)})^2 \,,\quad \hat G_{(3)}^i = \Sigma^2{\ast}_5F^i_{(2)} + F^i_{(2)}\wedge(d\theta +\sqrt{2}{A}^0_{(1)}) \,,
\end{equation}
where we put hats on the six-dimensional fields, define the $SO(1,1)$ scalar $\Sigma = e^{-\phi/\sqrt{6}}$, and $F^i_{(2)} = dA^i_{(1)}$, and have already imposed the anti-self-duality condition on $\hat G^i_{(3)}$. With this ans\"atze, we find that the equations of motion of the $\mc{N}=(4,0)$ theory are encoded in the five-dimensional Lagrangian, 
\begin{equation}
\mc{L}_5 = R{\ast}_51 - 3\Sigma^{-2}d\Sigma\wedge{\ast}_5d\Sigma - \Sigma^{-4}F^0_{(2)}\wedge{\ast}_5F^0_{(2)} - \Sigma^2F^i_{(2)}\wedge{\ast}_5F^i_{(2)} -\sqrt{2}A^0_{(1)}\wedge F^i_{(2)}\wedge F^i_{(2)} \,,
\end{equation} 
where $F^0_{(2)}=dA^0_{(2)}$. This is the canonical Lagrangian of the pure $\mc{N}=4$ supergravity theory in five dimensions \cite{Schon:2006kz}. We can now employ our embedding \eqref{M5embedding} to embed this $\mc{N}=4$ supergravity theory on the M5-brane worldvolume,
\begin{equation}\label{D5embeddingMtheory}
\begin{split}
&d\hat s^2_{11} = H^{-1/3}\left[\Sigma\,ds^2_5 + \Sigma^{-3}(d\theta + \sqrt{2}{A}^0_{(1)})^2\right] + H^{2/3}dy^idy^i \,,\\
&\hat F_{(4)} = \left[\Sigma^2{\ast}_5F^i_{(2)} + F^i_{(2)}\wedge(d\theta +\sqrt{2}{A}^0_{(1)})\right]\wedge dy^i - {\ast}_{\mb{R}^5}dH \,,\quad H = 1+\frac{\pi N}{r^3} \,.
\end{split}
\end{equation}
Here, ${\ast}_5$ is the Hodge dual with respect to the five-dimensional, worldvolume metric $ds^2_5$, and ${\ast}_{\mb{R}^5}$ is the Hodge dual with respect to the flat metric on the transverse $\mb{R}^5$. This can then be interpreted as the Type IIA solution\footnote{Our conventions for Type IIA supergravity and its relation to M-theory are presented in Appendix \ref{Mtheory}.}
\begin{equation}\label{D4embeddingIIA}
\begin{split}
&d\tilde s^2_{10} = H^{-3/8}\Sigma^{5/8}ds^2_5 + H^{5/8}\Sigma^{-3/8}dy^idy^i \,,\quad e^{\tilde\Phi} = H^{-1/4}\Sigma^{-9/4}\,,\\
&\tilde F_{(2)} = \sqrt{2}F^0_{(2)} \,,\quad \tilde H_{(3)} = -F^i_{(2)}\wedge dy^i \,,\\
&\tilde F_{(4)} = \Sigma^2{\ast}_5F^i_{(2)}\wedge dy^i - {\ast}_{\mb{R}^5}dH \,,
\end{split}
\end{equation}
which describes pure five-dimensional $\mc{N}=4$ supergravity embedded on the worldvolume of D4-branes. 

\subsection{Vertical example: $\text{M5}\to\text{NS5}$}

The vertical descendant of the M5-brane, \ie the NS5-brane, has a six-dimensional worldvolume. Thus, the worldvolume supergravity theory on the NS5-brane should be the same as the worldvolume supergravity of the M5-brane. Let $y^i = (y^a, \theta)$, $a\in\{1,\dots,4\}$, be the decomposition $\mb{R}^5=\mb{R}^4\times\mb{R}$ of the transverse space, and let $H$, the harmonic function on $\mb{R}^5$, depend only on $y^a$. This means that
\begin{equation}
H = 1 + \frac{N}{r^2} \,,\quad r^2 = y^ay^a \,.
\end{equation}
With this choice of harmonic function, the M5-brane solution with its worldvolume supergravity embedded becomes
\begin{equation}\label{NS5embeddingMtheory}
\begin{split}
&d\hat s^2_{11} = H^{-1/3}g_{\mu\nu}dx^\mu dx^\nu + H^{2/3}(dy^ady^a + d\theta^2) \,,\\
&\hat F_{(4)} = G^a_{(3)}\wedge dy^a + G^5_{(3)}\wedge d\theta - {\ast}_{4}dH\wedge d\theta \,,\quad H = 1+\frac{N}{r^2} \,,
\end{split}
\end{equation}
where ${\ast}_4$ is the Hodge dual with respect to the flat metric on $\mb{R}^4$, and we recall that $G^a_{(3)}$ and $G^5_{(3)}$ are anti-self-dual. This can then be interpreted as the Type IIA solution
\begin{equation}
\begin{split}
&d\tilde s^2_{10} = H^{-1/4}g_{\mu\nu}dx^\mu dx^\nu + H^{3/4}dy^ady^a \,,\quad e^{\tilde\Phi} = H^{1/2} \,,\\
&\tilde H_{(3)} = G^5_{(3)} \,,\\
&\tilde F_{(4)} = G^a_{(3)}\wedge dy^a - {\ast}_4dH \,,
\end{split}
\end{equation}
which describes the pure six-dimensional $\mc{N}=(4,0)$ supergravity theory embedded on the NS5-brane worldvolume. 

\section{Conclusion}

In this paper, we have demonstrated how noncompact Kaluza-Klein consistent reductions to corresponding supergravities can be made onto the worldvolumes of D3-branes in $D=10$ Type IIB supergravity and of M5-branes in $D=11$ M-theory or $D=10$ Type IIA supergravity. We also have showed how further diagonal or vertical dimensional reductions can be made consistently with the braneworld embeddings. These results clearly suggest a conjecture that consistent Kaluza-Klein reductions can be made onto the worldvolumes of any brane solutions with unbroken supersymmetry to worldvolume supergravity theories with the type of supersymmetry as the underlying \say{skeleton} brane. We have constructed the consistent-truncation Kaluza-Klein ans\"atze for the bosonic sectors of the illustrative cases studied, but extensions to include the corresponding fermionic fields are not expected to pose key difficulties.

Another question is how the braneworld supergravity embeddings that we have found relate to string theory constructions. Clearly, in a perturbative $\alpha'$ expansion, the general Weyl-anomaly beta function relations between string and gravitational field theories \cite{Callan:1985ia} gives a leading-order in $\alpha'$ consistency condition for closed strings on the embedded supergravity backgrounds we have found. Proceeding to all-orders in $\alpha'$ is more delicate. Replacement of flat-space worldvolume conformal blocks with conformal blocks of equivalent central charge or using string-theory dualities are approaches that have been used to generate all-orders consistent superstring backgrounds with \say{curved} worldvolumes \cite{Papadopoulos:1999tw}. Further string-theory implications of the supergravity-level embeddings we have found remain as interesting areas for further development.

\subsection*{Acknowledgments}

We would like to acknowledge helpful discussions with Matthew Cheung, Amit Sever, Arkady Tseytlin and Dan Waldram. The work of KSS was supported in part by the STFC under Consolidated Grant ST/P000762/1.

\begin{appendix}

\section{Type IIB supergravity conventions} \label{IIBeoms}

Our conventions are taken from \cite{Gauntlett:2010vu}. The bosonic sector of Type IIB supergravity contains the RR forms $\hat F_{(1)}$, $\hat F_{(3)}$, $\hat F_{(5)}$, the NSNS three-form $\hat H_{(3)}$, the dilaton $\hat\Phi$ and the metric, which will be taken to be the Einstein frame metric. The Bianchi identities are 
\begin{eqnarray}
&& d\hat F_{(5)} + \hat F_{(3)} \wedge  \hat H_{(3)} =0 \label{F5} \,,\\
&& d\hat F_{(3)}  + \hat F_{(1)} \wedge \hat H_{(3)}=0 \label{F3}\,, \\
&& d\hat H_{(3)} =0 \label{H3}\,,\\
&& d\hat F_{(1)} =0 \label{F1} \,,
\end{eqnarray}
which can be solved by introducing potentials as 
\begin{equation}
\begin{split}
&\hat F_{(1)} = d\hat C_0 \,,\quad \hat H_{(3)} = d\hat B_{(2)}\,,\\
&\hat F_{(3)} = d\hat C_{(2)} -\hat C_0  d\hat B_{(2)}\,,\quad \hat F_{(5)} = d\hat C_{(4)} -\hat C_{(2)} \wedge  \hat H_{(3)} \,.
\end{split}
\end{equation}
The equations of motion read
\begin{eqnarray}
&& \hat{\ast}\hat F_{(5)} =\hat F_{(5)} \label{eomF5} \,,\\
&& d(e^{\hat\Phi} \hat{\ast}\hat F_{(3)})  - \hat F_{(5)} \wedge \hat H_{(3)}=0 \label{eomF3}\,, \\
&& d(e^{-\hat\Phi} \hat{\ast}\hat H_{(3)})  -e^{\hat\Phi} \hat F_{(1)} \wedge \hat{\ast}\hat F_{(3)} -\hat F_{(3)} \wedge \hat F_{(5)}  =0 \label{eomH3} \,,\\
&& d(e^{2\hat\Phi} \hat{\ast}\hat F_{(1)})  +e^{\hat\Phi} \hat H_{(3)} \wedge \hat{\ast}\hat F_{(3)} =0 \label{eomF1} \,,\\
&& d\hat{\ast}d\hat\Phi -e^{2\hat\Phi} \hat F_{(1)} \wedge \hat{\ast}\hat F_{(1)} +\frac{1}{2}e^{-\hat\Phi} \hat H_{(3)} \wedge \hat{\ast}\hat H_{(3)} -\frac{1}{2} e^{\hat\Phi} \hat F_{(3)} \wedge \hat{\ast}\hat F_{(3)} =0 \label{eomPhi}\,,\\
&&\hat R_{MN} =  \frac{1}{2} \hat\nabla_M \hat\Phi \hat\nabla_N \hat\Phi+\frac{1}{2} e^{2\hat\Phi} \hat\nabla_M C_{0} \hat\nabla_N C_{0}
      + \frac{1}{96}\hat F^{}_{MP_1P_2P_3P_4}\hat F_N^{\ph{M}P_1P_2P_3P_4}  \nonumber \\ && \quad \qquad
      + \frac{1}{4}e^{-\hat\Phi}\left(
         \hat H^{}_{MP_1P_2} \hat H_{N}^{\ph{N}P_1P_2}
         - \frac{1}{12}(\hat H_{(3)})^2\hat g_{MN} \right) \nonumber \\ && \quad \qquad
      + \frac{1}{4}e^{\hat\Phi}\left(
         \hat F^{}_{MP_1P_2} \hat F_{N}^{\ph{N}P_1P_2}
         - \frac{1}{12}(\hat F_{(3)})^2\hat g_{MN}\right) \,. \label{IIBEinstein}
\end{eqnarray}
These admit a trombone symmetry, where 
\begin{equation}\label{tromboneIIB}
\begin{split}
&\hat g_{MN}\mapsto k^2 \hat g_{MN}\,,\quad \hat\Phi\mapsto \hat\Phi\,,\quad \hat C_0\mapsto \hat C_0\,,\\
& \hat F_{(3)}\mapsto k^2\hat F_{(3)}\,,\quad \hat H_{(3)}\mapsto k^2\hat H_{(3)}\,,\quad \hat F_{(5)}\mapsto k^4\hat F_{(5)} \,,
\end{split}
\end{equation}
for a constant $k$. 

In a configuration where $\hat\Phi = 0$ and $\hat F_{(1)} = 0$, it is convenient to package the three-form fluxes into a complex three-form
\begin{equation}
\hat G_{(3)} = \hat F_{(3)} + i\hat H_{(3)} \,,
\end{equation}
whose Bianchi identity and equation of motion are
\begin{equation}
d\hat G_{(3)} = 0 \,,
\end{equation}
and
\begin{equation}\label{eomG3}
d\hat {\ast}\hat G_{(3)} - i\hat G_{(3)}\wedge \hat F_{(5)} = 0 \,.
\end{equation}
The Bianchi identity for the five-form then reads
\begin{equation}\label{eomF5G3}
d\hat F_{(5)} + \frac{i}{2}\hat G_{(3)}\wedge\ov{\hat{G}}_{(3)} = 0 \,,
\end{equation}
and the consistency conditions for setting $\hat\Phi= 0$ and $\hat F_{(1)} = 0$ are contained in the single condition
\begin{equation}\label{compatibilityG3}
\hat G_{(3)}\wedge\hat{\ast}\hat G_{(3)} = 0 \,.
\end{equation}
With this condition, the Einstein equations simplify to
\begin{equation}
\hat R_{MN} = \frac{1}{96}\hat F^{}_{MP_1P_2P_3P_4}\hat F_N^{\ph{M}P_1P_2P_3P_4} + \frac{1}{8}\left(\hat G^{}_{MP_1P_2}\ov{\hat G}_N^{\ph{N}P_1P_2} + \ov {\hat G}^{}_{MP_1P_2} \hat G_N^{\ph{N}P_1P_2}\right) \,.
\end{equation}

\section{Useful identities (D3)}\label{usefulD3}

From \eqref{N=4ansatz}, we have
\begin{equation}
\hat H_{\mu\nu\Lambda} = \frac{1}{\sqrt{2}}\mc{F}^\Lambda_{\mu\nu}\,,\quad \hat F_{\mu\nu\Lambda} = -\frac{1}{\sqrt{2}}e^{-\phi}\tilde{\mc{F}}^\Lambda_{\mu\nu}\,,
\end{equation}
where $\tilde{\mc{F}}^\Lambda_{\mu\nu}$ are the components of ${\ast}_4\mc{F}^\Lambda_{(2)}$. Then,
\begin{equation}
\begin{split}
&\hat H^{}_{\mu P_1P_2}\hat H_\nu^{\ph{\nu}P_1P_2} = \mc{F}^\Lambda_{\mu\rho}\mc{F}_\nu^{\Lambda\,\rho} \,,\\
&\hat H^{}_{\mu P_1P_2}\hat H_\Sigma^{\ph{\Sigma}P_1P_2} = 0 \,,\\
&\hat H^{}_{\Lambda P_1P_2} \hat H_{\Sigma}^{\ph{\Sigma}P_1P_2} = \frac{1}{2}H \mc{F}^\Lambda_{\mu\nu}\mc{F}^{\Sigma\,\mu\nu} \,,
\end{split}
\end{equation}
and
\begin{equation}
\begin{split}
&\hat F^{}_{\mu P_1P_2}\hat F_\nu^{\ph{\nu}P_1P_2} = -\frac{1}{2}e^{-2\phi}(\mc{F}^\Lambda_{(2)})^2g_{\mu\nu} + e^{-2\phi}\mc{F}^\Lambda_{\mu\rho}\mc{F}_\nu^{\Lambda\,\rho} \,,\\
&\hat F^{}_{\mu P_1P_2}\hat F_\Sigma^{\ph{\Sigma}P_1P_2} = 0 \,,\\
&\hat F^{}_{\Lambda P_1P_2} \hat F_{\Sigma}^{\ph{\Sigma}P_1P_2} = -\frac{1}{2}e^{-2\phi}H \mc{F}^\Lambda_{\mu\nu}\mc{F}^{\Sigma\,\mu\nu} \,,
\end{split}
\end{equation}
where the indices on the RHS are raised by either $g^{\mu\nu}$ or $\delta^{\Lambda\Sigma}$.

\section{M-theory and Type IIA supergravity conventions} \label{Mtheory}

The bosonic sector of M-theory consists of a metric $\hat g_{MN}$ and a four-form flux $\hat F_{(4)}$. The Bianchi identity for the four-form is
\begin{equation}\label{bianchiF4}
d\hat F_{(4)} = 0 \,,
\end{equation}
and the equations of motion are
\begin{eqnarray}
&&d\hat{\ast}\hat F_{(4)} - \frac{1}{2}\hat F_{(4)}\wedge\hat F_{(4)} = 0 \,, \label{eomF411d} \\
&&\hat R_{MN} = \frac{1}{12}\left(\hat F_{MPQR}\hat F_N^{\ph{N}PQR} - \frac{1}{12}(\hat F_{(4)})^2\hat g_{MN}\right) \,.\label{11einstein}
\end{eqnarray}
The dynamics of Type IIA supergravity are encoded in M-theory about backgrounds where the eleven-dimensional spacetime has an $U(1)$ isometry. We write
\begin{equation}
\begin{split}
&d\hat s^2_{11} = e^{-\tilde\Phi/6}d\tilde s^2_{10}  +e^{4\tilde\Phi/3}(d\theta + \tilde A_{(1)})^2 \,,\\
&\hat F_{(4)} = \tilde F_{(4)} + \tilde H_{(3)}\wedge (d\theta + \tilde A_{(1)}) \,,
\end{split}
\end{equation}
where the ten-dimensional Type IIA fields are distinguished with a tilde and are independent of $\theta$. The metric $\tilde g$ is the Einstein frame metric of the Type IIA theory. The Bianchi identity of $\hat F_{(4)}$ yields
\begin{equation}
d\tilde H_{(3)} = 0 \,,\quad d\tilde F_{(4)} - \tilde H_{(3)}\wedge \tilde F_{(2)} = 0 \,,
\end{equation}
where $\tilde F_{(2)} = d\tilde A_{(1)}$. These can be integrated by introducing the NSNS two-form potential $\tilde B_{(2)}$ and RR three-form potential $\tilde A_{(3)}$,
\begin{equation}
\tilde H_{(3)} = d\tilde B_{(2)} \,,\quad \tilde F_{(4)} = d\tilde A_{(3)} - \tilde H_{(3)}\wedge \tilde A_{(1)} \,.
\end{equation}

\section{Useful identities (M5)}\label{usefulM5}

From \eqref{M5embedding}, the non-zero components of $\hat F_{(4)}$ are
\begin{equation}
\hat F_{\mu\nu\rho i}= G^i_{\mu\nu\rho}\,,\quad \hat F_{ijkl} = -\epsilon^p_{\ph{p}ijkl}\partial_pH \,,
\end{equation}
where the $i,j,\dots$ indices are raised by $\delta^{ij}$. Then,
\begin{equation}
\begin{split}
&\hat F_{\mu P_1P_2P_3}\hat F_\nu^{\ph{\nu}P_1P_2P_3} = 3G^i_{\mu\sigma_1\sigma_2}G_\nu^{i\,\sigma_1\sigma_2} \,,\\
&\hat F_{\mu P_1P_2P_3}\hat F_i^{\ph{i}P_1P_2P_3} = 0 \,,\\
&\hat F_{iP_1P_2P_3}\hat F_j^{\ph{j}P_1P_2P_3} = 6H^{-2}\left(\partial_kH\partial^kH\delta_{ij} - \partial_iH\partial_jH\right) \,,
\end{split}
\end{equation}
where the indices on the RHS are raised by either $g^{\mu\nu}$ or $\delta^{ij}$. For the $(i,j)$ components, we used the anti-self-duality of $G^i_{(3)}$, which states that $G^i_{(3)}\wedge G^j_{(3)} = -G^i_{(3)}\wedge{\ast}_6G^j_{(3)} = G^j_{(3)}\wedge{\ast}_6G^i_{(3)} = G^j_{(3)}\wedge G^i_{(3)} = 0$.

\addcontentsline{toc}{section}{References}

\bibliographystyle{utphys}
\bibliography{sugra_on_branes}{}

\end{appendix}

\end{document}